\documentstyle[twoside]{article}

\catcode`\@=11
\long\def\@makefntext#1{
\protect\noindent \hbox to 3.2pt {\hskip-.9pt

$^{{\eightrm\@thefnmark}}$\hfil}#1\hfill}               

\def\thefootnote{\fnsymbol{footnote}}
\def\@makefnmark{\hbox to 0pt{$^{\@thefnmark}$\hss}}    

\def\ps@myheadings{\let\@mkboth\@gobbletwo
\def\@oddhead{\hbox{}
\rightmark\hfil\eightrm\thepage}

\def\@oddfoot{}\def\@evenhead{\eightrm\thepage\hfil
\leftmark\hbox{}}\def\@evenfoot{}
\def\sectionmark##1{}\def\subsectionmark##1{}}

\oddsidemargin=\evensidemargin
\renewcommand{\thefootnote}{\fnsymbol{footnote}}

\newcounter{sectionc}\newcounter{subsectionc}\newcounter{subsubsectionc}
\renewcommand{\section}[1] {\vspace{12pt}\addtocounter{sectionc}{1}

\setcounter{subsectionc}{0}\setcounter{subsubsectionc}{0}\noindent

        {\tenbf\thesectionc. #1}\par\vspace{5pt}}
\renewcommand{\subsection}[1] {\vspace{12pt}\addtocounter{subsectionc}{1}

        \setcounter{subsubsectionc}{0}\noindent

        {\bf\thesectionc.\thesubsectionc. {\kern1pt \bfit #1}}\par\vspace{5pt}}
\renewcommand{\subsubsection}[1] {\vspace{12pt}\addtocounter{subsubsectionc}{1}
        \noindent{\tenrm\thesectionc.\thesubsectionc.\thesubsubsectionc.
        {\kern1pt \tenit #1}}\par\vspace{5pt}}
\newcommand{\nonumsection}[1] {\vspace{12pt}\noindent{\tenbf #1}
        \par\vspace{5pt}}

\newcounter{appendixc}
\newcounter{subappendixc}[appendixc]
\newcounter{subsubappendixc}[subappendixc]
\renewcommand{\thesubappendixc}{\Alph{appendixc}.\arabic{subappendixc}}
\renewcommand{\thesubsubappendixc}
        {\Alph{appendixc}.\arabic{subappendixc}.\arabic{subsubappendixc}}

\renewcommand{\appendix}[1] {\vspace{12pt}
        \refstepcounter{appendixc}
        \setcounter{figure}{0}
        \setcounter{table}{0}
        \setcounter{lemma}{0}
        \setcounter{theorem}{0}
        \setcounter{corollary}{0}
        \setcounter{definition}{0}
        \setcounter{equation}{0}
        \renewcommand{\thefigure}{\Alph{appendixc}.\arabic{figure}}
        \renewcommand{\thetable}{\Alph{appendixc}.\arabic{table}}
        \renewcommand{\theappendixc}{\Alph{appendixc}}
        \renewcommand{\thelemma}{\Alph{appendixc}.\arabic{lemma}}
        \renewcommand{\thetheorem}{\Alph{appendixc}.\arabic{theorem}}
        \renewcommand{\thedefinition}{\Alph{appendixc}.\arabic{definition}}
        \renewcommand{\thecorollary}{\Alph{appendixc}.\arabic{corollary}}
        \renewcommand{\theequation}{\Alph{appendixc}.\arabic{equation}}
        \noindent{\tenbf Appendix \theappendixc #1}\par\vspace{5pt}}
\newcommand{\subappendix}[1] {\vspace{12pt}
        \refstepcounter{subappendixc}
        \noindent{\bf Appendix \thesubappendixc. {\kern1pt \bfit #1}}
        \par\vspace{5pt}}
\newcommand{\subsubappendix}[1] {\vspace{12pt}
        \refstepcounter{subsubappendixc}
        \noindent{\rm Appendix \thesubsubappendixc. {\kern1pt \tenit #1}}
        \par\vspace{5pt}}

\newcommand{\textlineskip}{\baselineskip=13pt}
\newcommand{\smalllineskip}{\baselineskip=10pt}

\def\eightcirc{
\begin{picture}(0,0)
\put(4.4,1.8){\circle{6.5}}
\end{picture}}
\def\eightcopyright{\eightcirc\kern2.7pt\hbox{\eightrm c}}

\def\abstracts#1#2#3{{
        \centering{\begin{minipage}{4.5in}\baselineskip=10pt\footnotesize
        \parindent=0pt #1\par

        \parindent=15pt #2\par
        \parindent=15pt #3
        \end{minipage}}\par}}

\newcommand{\bibit}{\nineit}

\renewenvironment{thebibliography}[1]
        {\frenchspacing
         \ninerm\baselineskip=11pt
         \begin{list}{\arabic{enumi}.}
        {\usecounter{enumi}\setlength{\parsep}{0pt}
         \setlength{\leftmargin 12.7pt}{\rightmargin 0pt} 
         \setlength{\itemsep}{0pt} \settowidth
        {\labelwidth}{#1.}\sloppy}}{\end{list}}

\newcounter{itemlistc}
\newcounter{romanlistc}
\newcounter{alphlistc}
\newcounter{arabiclistc}

\newcommand{\fcaption}[1]{
        \refstepcounter{figure}
        \setbox\@tempboxa = \hbox{\footnotesize Fig.~\thefigure. #1}
        \ifdim \wd\@tempboxa > 5in
           {\begin{center}
        \parbox{5in}{\footnotesize\smalllineskip Fig.~\thefigure. #1}
            \end{center}}
        \else
             {\begin{center}
             {\footnotesize Fig.~\thefigure. #1}
              \end{center}}
        \fi}

\newcommand{\tcaption}[1]{
        \refstepcounter{table}
        \setbox\@tempboxa = \hbox{\footnotesize Table~\thetable. #1}
        \ifdim \wd\@tempboxa > 5in
           {\begin{center}
        \parbox{5in}{\footnotesize\smalllineskip Table~\thetable. #1}
            \end{center}}
        \else
             {\begin{center}
             {\footnotesize Table~\thetable. #1}
              \end{center}}
        \fi}

\def\@citex[#1]#2{\if@filesw\immediate\write\@auxout
        {\string\citation{#2}}\fi
\def\@citea{}\@cite{\@for\@citeb:=#2\do
        {\@citea\def\@citea{,}\@ifundefined
        {b@\@citeb}{{\bf ?}\@warning
        {Citation `\@citeb' on page \thepage \space undefined}}
        {\csname b@\@citeb\endcsname}}}{#1}}

\newif\if@cghi
\def\cite{\@cghitrue\@ifnextchar [{\@tempswatrue
        \@citex}{\@tempswafalse\@citex[]}}
\def\citelow{\@cghifalse\@ifnextchar [{\@tempswatrue
        \@citex}{\@tempswafalse\@citex[]}}
\def\@cite#1#2{{$\null^{#1}$\if@tempswa\typeout
        {IJCGA warning: optional citation argument

        ignored: `#2'} \fi}}

\def\pmb#1{\setbox0=\hbox{#1}
        \kern-.025em\copy0\kern-\wd0
        \kern.05em\copy0\kern-\wd0
        \kern-.025em\raise.0433em\box0}

\def\fnt#1#2{\footnotetext{\kern-.3em
        {$^{\mbox{\scriptsize #1}}$}{#2}}}

\def\fpage#1{\begingroup
\voffset=.3in
\thispagestyle{empty}\begin{table}[b]\centerline{\footnotesize #1}
        \end{table}\endgroup}

\def\runninghead#1#2{\pagestyle{myheadings}
\markboth{{\protect\footnotesize\it{\quad #1}}\hfill}
{\hfill{\protect\footnotesize\it{#2\quad}}}}
\font\tenrm=cmr10
\font\tenit=cmti10

\font\tenbf=cmbx10
\font\bfit=cmbxti10 at 10pt
\font\ninerm=cmr9
\font\nineit=cmti9

\font\eightrm=cmr8

\def\qed{\hbox{${\vcenter{\vbox{                        
   \hrule height 0.4pt\hbox{\vrule width 0.4pt height 6pt
   \kern5pt\vrule width 0.4pt}\hrule height 0.4pt}}}$}}

\renewcommand{\thefootnote}{\fnsymbol{footnote}}        

\begin{document}

\runninghead{D. Cangemi}{Self-Duality and Maximally Helicity Violating QCD
  Amplitudes}

\normalsize\textlineskip
\thispagestyle{empty}
\setcounter{page}{1}

\vspace*{-2.5cm}\smalllineskip{\hbox{}\hfill{NBI-HE-96-51}}

\vspace*{0.88truein}

\fpage{1} \centerline{\bf SELF-DUALITY AND MAXIMALLY HELICITY
  VIOLATING}
\vspace*{0.035truein}
\centerline{\bf QCD AMPLITUDES}
\vspace*{0.37truein}
\centerline{\footnotesize Daniel CANGEMI}
\vspace*{0.015truein}
\centerline{\footnotesize\it Niels Bohr Institute, Blegdamsvej 17}
\baselineskip=10pt
\centerline{\footnotesize\it DK -- 2500 Copenhagen \O, Denmark.}
\vspace*{0.225truein}
\centerline{September 1996}

\vspace*{0.21truein}
\abstracts{ I review some recent work that describes the close analogy
  between self-dual Yang--Mills amplitudes and QCD amplitudes with
  external gluons of positive helicity. This analogy is carried at
  tree level for amplitudes with two external quarks and up to
  one-loop for amplitudes involving only external gluons.
}{}{}

\textheight=7.8truein
\setcounter{footnote}{0}
\renewcommand{\thefootnote}{\alph{footnote}}

\section{Introduction}
\noindent
Low-dimensional field theories are extensively investigated due to
their greater tractability. Although the mathematical structure
emerging from lower dimensional models is often worth some attention
in itself, one really hopes to learn something about the physical world.
This has often been the case: two-dimensional QED and QCD have taught
us a lot on asymptotic freedom, three-dimensional Chern-Simons theory
has played a crucial r\^ole in planar condensed matter systems.

Low-dimensional field theories are sometimes completely integrable.
Moreover all known integrable models derive from the self-dual
Yang--Mills (SDYM) equations. I will present in this article a
physical application of these equations and show an explicit relation
between a solution of the SDYM equations and some QCD amplitudes.  The
amplitudes that I will discuss involve external gluons with all the
same helicity. Since these amplitudes are known to vanish at tree
level, they are referred to as Maximally Helicity Violating (MHV)
amplitudes.  The notions of self-duality and positive helicity
coincide for Maxwell (free) fields and are intimately connected in the
case of non-Abelian gauge fields.\cite{duff} A link between MHV
amplitudes and integrable models was pointed out by Nair.\cite{nair}
Recently, Bardeen\cite{bardeen,selivanov} has shown the direct
relevance of a SDYM solution in the calculation of MHV tree
amplitudes. This analysis has then been extended to one-loop
amplitudes.\cite{daniel,gordon}

The probe of physics beyond the Standard Model needs more and more
accurate data, not only from experimentalists but also from
theoreticians. The leading order of $\alpha_s$ is obtained by squaring
tree-level scattering amplitudes but the next-to-leading order is also
necessary. To this end people have developed ingenious methods to
calculate tree and one-loop QCD amplitudes.\cite{zvi-review} After
lengthy and complex calculations, some of these amplitudes turn out to
be very simple. Tree amplitudes with all or all but one external
gluons with positive helicity\footnote{By convention, helicity is
  defined for outgoing particles.}\ \ are zero,
\begin{eqnarray}
  \label{tree-ampl}
  A^{\rm tree} _n ( g_1^+ , \ldots , g_n^+ ) = 0 \,, \\
  A^{\rm tree} _n ( g_1^- , g_2^+ , \ldots , g_n^+ ) = 0 \,.
\end{eqnarray}
Non trivial tree amplitudes\footnote{Only the color-leading
  partial amplitudes are shown here. I also use the spinor notation
  where $\langle 1 2 \rangle = - ( k_{1,0+z} / k_{1,x+iy} - k_{2,0+z}
  / k_{2,x+iy} ) k_{1,x+iy} k_{2,x+iy} / \sqrt{|k_{1,0+z}k_{2,0+z}|}$
  and $[ 1 2 ] = {\rm sign} ( k_{1,0} k_{2,0} ) \langle 2 1
  \rangle^*$. Details on color ordering and the spinor notation can be
  found elsewhere.\cite{zvi-review,daniel}}\ \ are for example the
Park-Taylor\cite{parke} amplitudes with two negative helicity external
gluons,
\begin{equation}
  A^{\rm tree} _n ( g_1^- , g_2^- , g_3^+ , \ldots , g_n^+ ) =  i
    g^4 \frac{ \langle 1 2 \rangle^4}{ \langle 1 2 \rangle \langle 2 3
    \rangle \cdots \langle n 1 \rangle } \,. \\
\end{equation}
Whereas amplitudes~(\ref{tree-ampl}) vanish at tree level, their
one-loop correction is non zero and has a very simple structure. The
four-point function is:\cite{kosower}
\begin{equation}
  \label{one-loop}
  A^{\rm one-loop}_4 ( g_1^+ , g_2^+, g_3^+ , g_4^+ ) = - N_c \; g^4
  \frac{i}{48 \pi^2} \frac{ [ 1 2 ] [ 3 4 ] }{ \langle 1 2 \rangle
  \langle 3 4 \rangle } \,,
\end{equation}
and their form is explicitly known for an arbitrary number of external
gluons.\cite{bcdkm}

In view of these results, several questions arise naturally. Their
answers will be the object of this article:
\begin{enumerate}
\item Why is the form of these amplitudes so simple?
\item Is there an ``effective model'' for QCD restricted to positive
  helicity configurations which reproduces these amplitudes?
\item Is there a symmetry behind the vanishing of the tree
  amplitude~(\ref{tree-ampl}) ?
\item Is the one-loop result~(\ref{one-loop}) a manifestation of an
  anomaly?
\end{enumerate}
At the present time, the few answers and the possible hints that we
have are listed hereafter and will be developed in the following
sections.
\begin{enumerate}
\item It is clear that helicity is the relevant factor. A positive
  helicity configuration is self-dual: a positive helicity
  electromagnetic wave has $\vec E = i \vec B$, which is nothing else
  that the self-dual equation $F_{\mu\nu} = i
  \epsilon_{\mu\nu\rho\sigma} F^{\rho\sigma}$.
\item Nair first showed\cite{nair} that the amplitude $A^{\rm
    tree}_n(g_1^-, g_2^-, g_3^+, \ldots, g_n^+)$ can be derived from
  the current algebra of a $k=1$ WZNW model based on $CP^1$.  Keeping
  with the analogy between positive helicity and self-duality, Bardeen
  observed\cite{bardeen} that some solutions of the SDYM equations
  reproduce the QCD amplitudes $A^{\rm tree}_n(g_1^+, \ldots, g_n^+)$.
  This observation can be extended\cite{daniel,gordon} to the one-loop
  amplitudes~(\ref{one-loop}) using various quantizations of the SDYM
  equations. One of them\cite{gordon} is directly related to the
  self-dual sector of QCD, a truncation of $N=4$ SYM theory in the
  light-cone formalism. It is interesting to notice that one of the
  other quantizations involves an action which is just a
  generalization\cite{moore} in four dimensions of the WZNW action, a
  possible link to the earlier work of Nair.
\item Up to now, the identification of the self-dual amplitudes with
  the MHV amplitudes has in no way simplified the actual calculations.
  One would like a deeper understanding of the vanishing of the tree
  amplitudes, for example using some symmetry argument. Indeed, the
  SDYM equations possess an infinite symmetry that forms an affine Lie
  algebra and it is suspected, although not proven, that this large
  symmetry is responsible for the vanishing of the tree amplitudes.
\item Assuming that the previous argument holds at tree level, one
  concludes that the one-loop results~(\ref{one-loop}) should be a
  consequence of an anomaly.
\end{enumerate}
These are the various pieces of the puzzle and we will now see how
they fit together.

After this introduction, I review the SDYM equations, their solutions
and their symmetries. I discuss the tree amplitudes in section 3 and
the one-loop amplitudes in section 4. I conclude with some remarks in
section 5.

\section{SDYM equations}
\noindent
SDYM equations have real solutions only in Euclidean or
$(2+2)$-signature spacetime. We here work in Minkowski spacetime,
where they are:
\begin{equation}
  \label{SDYM}
  F_{\mu\nu} = \frac{i}{2} \epsilon_{\mu\nu\rho\sigma}
  F^{\rho\sigma} \,,
\end{equation}
and have complex solutions.  It was remarked by Duff and
Isham\cite{duff} that complex field configurations are naturally
provided by off-diagonal matrix elements of the hermitian field
operator. In the Yang-Mills case, they further showed that a solution
to the equations of motion with definite duality is given by a matrix
element between the vacuum and a coherent state with definite
helicity. We shall come back to this point in the next section. In
com\-po\-nents,\footnote{I use the conventions $k_{0\pm z} = k_0 \pm
  k_z$ and $k_{x\pm i y} = k_x \pm i k_y$, with metric $g_{\mu\nu} =
  {\rm diag}(1,-1,-1,-1)$ and totally antisymmetric tensor
  $\epsilon_{0123} = 1$.}\ \ the SDYM equations read:
\begin{equation}
  \label{sdym}
  F_{0+z,x-i y} = 0 \,, \qquad
  F_{0-z,x+i y} = 0 \,, \qquad
  F_{0-z,0+z} = F_{x+i y,x-i y} \,.
\end{equation}
Dimensional reductions of these equations yield various (maybe even
all\cite{ward}) integrable models. The $(2+1)$-dimensional
Chern-Simons system coupled to self-dual matter and discussed
elsewhere in these proceedings is such an example.

In Yang's approach,\cite{yang} the two first equations are taken
as two zero curvature conditions solved by:
\begin{eqnarray}
  &-\frac{i g}{\sqrt{2}} A_{0+z} = h^{-1} \partial_{0+z} h \,, \qquad
   -\frac{i g}{\sqrt{2}} A_{x-i y} =
     h^{-1} \partial_{x-i y} h \,,
  \nonumber \\[-6pt]
  \\[-6pt]
  &-\frac{i g}{\sqrt{2}} A_{0-z} =
     \bar h^{-1} \partial_{0-z} \bar h \,, \qquad
  -\frac{i g}{\sqrt{2}} A_{x+i y} =
     \bar h^{-1} \partial_{x+i y} \bar h \,. \nonumber
\end{eqnarray}
The last SDYM equation gives then:
\begin{equation}
  \partial_{0-z} ( H^{-1} \partial_{0+z} H ) -
  \partial_{x+i y} ( H^{-1} \partial_{x-i y} H ) = 0
  \,,
  \qquad H = h \bar h^{-1} \,.
\end{equation}
This resembles a two-dimensional conserved current equation and is in fact
obtained from an action similar to the WZNW action, first proposed by
Donaldson\cite{donaldson} and by Nair and Schiff,\cite{schiff}
\begin{eqnarray}
  \label{actionJ}
  S_{\rm DNS}(H) &=& \frac{f_\pi^2}{2} \int d^4x \;{\rm tr}\;
  \left( \partial_{0+z} H \partial_{0-z} H^{-1} -
    \partial_{x-i y} H \partial_{x+i y} H^{-1} \right)
  \nonumber\\
  &&{} + \frac{f_\pi^2}{2} \int d^4x dt
    \;{\rm tr}\; \left( \, [ H^{-1} \partial_{0+z} H, H^{-1}
    \partial_{0-z} H ] - \right.\nonumber\\
  &&{} \hspace{3cm} \left. [ H^{-1} \partial_{x-i y} H, H^{-1}
    \partial_{x+i y} H ] \; \right)  H^{-1} \partial_t H \,.
\end{eqnarray}
One checks that its $\beta$-function vanishes at one-loop and possibly
at all order.\cite{ketov}

Another way to solve the SDYM equations is to go in the light-cone
gauge $A_{0-z} = 0$, where the two last SDYM equations are solved by:
\begin{equation}
  \label{A-Phi-a}
  A_{x+i y} = 0 \,, \qquad A_{0+z} = \sqrt{2} \;
  \partial_{x+i y} \Phi \,, \qquad A_{x-i y} =
  \sqrt{2} \; \partial_{0-z} \Phi \,,
\end{equation}
in terms of a zero dimensional scalar field $\Phi$. The remaining SDYM
equation gives:
\begin{equation}
  \label{equPhi}
  \partial^2 \Phi - i  g \; [ \partial_{x+i y} \Phi,
  \partial_{0-z} \Phi ] = 0 \,.
\end{equation}
There are two actions associated to these equations of motion.

The first one was proposed by Leznov, Mukhtarov and Parkes.\cite{lmp} It does
not introduce new field,
\begin{equation}
  \label{actionPhi}
  S_{\rm LPM}(\phi) = f_\pi^2 \int d^4x \;{\rm tr}\;
  \left( \frac{1}{2} \partial \phi \cdot \partial \phi +
    \frac{i  g}{3} \; \phi \; [ \partial_{x+i y} \phi,
    \partial_{0-z} \phi ] \right) \,,
\end{equation}
but is not real, explicitly breaks Lorenz invariance and has a non
renormalizable interaction by power counting.

The second action, proposed by Chalmers and Siegel\cite{gordon},
uses a Lagrange multiplier of dimension two to enforce
Eq.~(\ref{equPhi}),
\begin{equation}
  \label{siegel-action}
  S_{\rm SC}(\phi,\Lambda) = - \int d^4x \; {\rm tr} \; \Lambda \left(
  \partial^2 \Phi -  i  g \; [ \partial_{x+i y} \Phi, \partial_{0-z} \Phi
  ] \right)
\end{equation}
This action is also obtained after a truncation of $N=4$ SYM theory in
light-cone formalism\cite{mandelstam,siegel} ($\Lambda$ and $\Phi$ correspond
then to the highest and lowest components of the $N=4$ chiral superfield). The
auxiliary field $\Lambda$ obeys the additional equation of motion:
\begin{equation}
  \label{equLambda}
  \partial^2 \Lambda - i  g \; [ \partial_{x+i y} \Lambda,
  \partial_{0-z} \Phi ] - i  g \; [ \partial_{x+i y} \Phi,
  \partial_{0-z} \Lambda ] = 0 \,,
\end{equation}
and appears only in tree amplitudes. In fact, it is impossible to draw
connected diagrams with more than one loop and thus this model only
admits tree and one-loop corrections! However the Hamiltonian is not
bounded by below so one should be careful in the full quantization of
the model.

The symmetries $\Phi \to \Phi + \Lambda / f_\pi^2$ of the SDYM equations in
their light-cone gauge form~(\ref{equPhi}) are also described by
Eq.~(\ref{equLambda}).  A one-parameter family of symmetries $\Lambda_s$ is
constructed from the following pair of recursion relations:\cite{dolan}
\begin{eqnarray}
  \label{Lambdas}
  &\partial_{0-z} \Lambda_{s+1} = \partial_{x-i y} \Lambda_s -
  i  g [ \partial_{0-z} \Phi, \Lambda_s ] \,,
  \nonumber\\[-6pt]
  \label{hierarchy} \\[-6pt]
  &\partial_{x+i y} \Lambda_{s+1} = \partial_{0+z} \Lambda_s -
  i  g [ \partial_{x+i y} \Phi, \Lambda_s ]
  \,. \nonumber
\end{eqnarray}
These equations are compatible if $\Lambda_s$ is a solution
of~(\ref{equLambda}). Moreover $\Lambda_{s+1}$ is a symmetry if $\Phi$ is a
solution of the SDYM equations. These symmetries are known to form an affine
Lie algebra. Of course, the $\Lambda$'s are symmetries of the equations of
motion and not necessarily of the action. Among all the $\Lambda_s$, only
$\Lambda_0 = T^a$ and $\Lambda_1 = -i g \; [ \phi, T^a ]$ are true symmetries
of the action.  Nevertheless, the hierarchy~(\ref{hierarchy}) defines an
infinite set of conserved currents whose classical expressions are:
\begin{eqnarray}
  &{\cal J}_{s,0+z} = \partial_{0+z} \Lambda_s - i g\; [
  \partial_{x+i y} \Phi, \Lambda_s ] \,, \quad
  {\cal J}_{s,x-i y} = \partial_{x-i y} \Lambda_s -
  i g\; [ \partial_{0-z} \Phi, \Lambda_s ] \,, \nonumber\\
  &\partial_\mu {\cal J}^\mu_s = \frac{1}{2} i g\; [ \Lambda_{s-1},
  \partial^2 \Phi - i g\; [ \partial_{x+i y} \Phi,
  \partial_{0-z} \Phi ] = 0 \,.
\end{eqnarray}
Since these currents are only known for classical solutions $\Phi$,
one can only derive tree level identities and not true Ward identities
relating Green functions of different orders in $\hbar$.

\section{Tree amplitudes}
\noindent
We now compare the tree amplitudes of these SDYM models with the QCD
ones. Since the three SDYM actions have equivalent equations of motion
they have equivalent tree amplitudes. Namely, there is a direct
relation between tree amplitudes and classical solutions.  An
amplitude is the on-shell truncation of a connected Green function. A
connected tree Green function is generated by the Legendre transform
$W(J)$ of the classical action $S(\phi)$,
\begin{eqnarray}
  & \displaystyle
  \frac{\delta S(\phi)}{\delta \phi(x)} \biggr|_{\phi=\Phi_J}
      + J(x) = 0 \,, \\
  & W(J) = S( \Phi_J ) + \int dx \; \Phi_J(x) J(x) \,,\\
  & \displaystyle
  \langle \phi(x_1) \cdots \phi(x_{n+1}) \rangle_c = \frac{ i
      \delta^{n+1} W(J) }{ i  \delta J(x_1) \cdots i
      \delta J(x_{n+1}) } \; \biggr|_{J=0} \,.
\end{eqnarray}
Since the classical solution $\Phi_J$ in presence of a source $J$ is
given by the first variation of the generating functional $W(J)$, we
have:
\begin{equation}
  \label{funct}
  \langle \phi(x_1) \cdots \phi(x_{n+1}) \rangle_c = \frac{ \delta^{n}
  \Phi_J(x_{n+1}) }{ i  \delta J(x_1) \cdots i   \delta
  J(x_n) } \; \biggr|_{J=0}  \,.
\end{equation}
The classical solution $\Phi_J$ is an infinite series in $J$ whose
coefficients are the connected Green functions.  In our case, the
equations of motion with source read:
\begin{equation}
  \label{equPhiJ}
  \partial^2 \Phi_J -  i  g \; [ \partial_{x+ i y} \Phi_J,
  \partial_{0-z} \Phi_J ] + J = 0
\end{equation}
whose solution is obtained by inverting the differential operator
$\partial^2$ with the Feynman propagator $-1/(k^2 +
i\epsilon)$. 

\newcommand{\jj}{\rlap{\~\ }j }
Following Duff and Isham,\cite{duff} let us consider a coherent
in-state (the minus sign is introduced to match earlier
conventions\cite{daniel}) $|\rlap{\~\ } j \rangle = \sum_{m=0}^\infty
\frac{1}{m!} | (- j)^m \rangle$ based on a positive energy on-shell
wave,
\begin{equation}
\label{planew}
  j(x) = - i \sum_{j=1}^n \; T^{a_j} e^{- i  k_j x} f(k_j) \,,
\end{equation}
(the function $f(k)$ has support on the light-cone). By LSZ reduction,
we construct the connected tree matrix element of the interacting
field operator $\hat\phi$ between this state and the vacuum,
\begin{eqnarray}
  \label{matrix-el}
  \lefteqn{
  \langle0| \hat\phi(k) |\rlap{\~\ } j\rangle^{\rm tree}_c =
  \sum_{m=1}^\infty \frac{i^m}{m!} \int \frac{d^4p_1}{(2\pi)^4} \cdots
  \frac{d^4p_m}{(2\pi)^4} } \qquad\qquad\qquad \nonumber\\
  &&{} \times p_1^2 \, j(-p_1) \cdots p_m^2 \, j(-p_m) \; \langle
  \phi(p_1) \cdots \phi(p_m) \phi(k) \rangle_c^{\rm tree} \,.
\end{eqnarray}
We are interested in graphs with a given number of external leg, say
$n+1$:
\begin{eqnarray}
  \label{currampl}
  \Bigl\langle \phi(k) \Bigr\rangle_{1 \cdots n} \equiv
  (-  i  k_1^2) f(k_1) \cdots (-  i  k_n^2) f(k_n)
  \left. \Bigl\langle \phi(k_1) \cdots \phi(k_n) \phi(k) \Bigr\rangle
  \right|_{k_1^2 = \cdots = k_n^2 = 0} \,. \nonumber\\
\end{eqnarray}
They are readily obtained from the matrix element~(\ref{matrix-el}).
One inserts~(\ref{planew}) and keeps the terms proportional to
$g^{n-1} f(k_1) \cdots f(k_n)$. A tree amplitude with $(n+1)$ external
legs is then simply $( - i k^2 ) \; \left. \bigl\langle \phi(k)
  \bigr\rangle_{1 \cdots n} \right|_{k^2=0}$.

The matrix element~(\ref{matrix-el}) is connected to the solution
$\Phi_J$ through the following identity:\cite{duff}
\begin{equation}
  \langle0| \hat\phi(k) |\rlap{\~\ } j\rangle^{\rm tree}_c = \exp \left[
  \int \frac{d^4p}{(2\pi)^4} \; p^2 j(p) \frac{\delta}{\delta J(-p)}
  \right] \Phi_J(k) \biggr|_{J=0} = \Phi_{J(p) = p^2 j(p)}(k) \,.
\end{equation}
We need thus to find a solution to the classical equations of motion with
an on-shell source $J(p) = p^2 j(p)$.  The solution is written as a
series in the coupling constant $g$,
\begin{equation}
  \Phi_J(x) = \sum_{m=1}^\infty \Phi_J^{(m)}(x) \,, \qquad
  \Phi_J^{(m)} \propto g^{m-1} \,.
\end{equation}
Using the explicit form of $j(p)$, we prove by
recurrence:\cite{bardeen,daniel}
\begin{eqnarray}
 \label{treePhi}
  \Phi^{(n)}(x) &=& -  i  g^{n-1} \sum_{{\rm permutations}
    \atop {\rm of\ }( 1 \cdots n )} T^{a_1} \cdots
  T^{a_n} \; e^{- i  (k_1 + \cdots + k_n)x} f(k_1) \cdots
  f(k_n) \nonumber\\
  &\times& (Q_1 - Q_2)^{-1} (Q_2 - Q_3)^{-1} \cdots (Q_{n-1} - Q_n)^{-1}
    \; + \cdots \,.
\end{eqnarray}
with $Q_i = k_{i,x+iy} / k_{i,0+z}$.  The piece written here involves
all momenta $k_j$ ($j = 1, \ldots, n$) and, as mentioned before, is
equal to the current amplitude~(\ref{currampl}).  The additional terms
denoted by the triple dots (as well as the other $\Phi^{(m)}, \;
m\not=n$) correspond to tree diagrams having two or more legs with the
same momentum.\cite{daniel,oota}

The remarkable property of this solution is that it does not involve
the multiparticle poles appearing in the intermediate states. In
particular there is no pole in $k^2 = (k_1 + \cdots + k_n)^2$ so that
the tree amplitudes (obtained by multiplying~(\ref{currampl}) by $k^2$
and taking the limit $k^2\to0$) vanish.

The MHV tree amplitudes~(\ref{tree-ampl}) vanish in exactly the same
way.  The Berends and Giele relations\cite{berends} derived in QCD
coincide for MHV configurations with the recurrence
relations obtained here. For $f(k) = Q / k_{x+ i y}$, we get the
following identity\cite{daniel} between the tree SDYM
current~(\ref{currampl}) and the corresponding tree MHV current in the
light-cone gauge as calculated for example by Mahlon {\it et
  al.}:\cite{mahlon}
\begin{eqnarray}
  \label{tree-id}
  \begin{array}{ll}
    \Bigl\langle A_{0+z}(k) \Bigr\rangle_{1^+ \cdots n^+} \; \rlap{\ \
      $=$}\raisebox{1em}{tree} \; -  i  \sqrt{2} \;
    k_{x+ i y} \; \Bigl\langle \phi(k) \Bigr\rangle_{1 \cdots n} \,, &
    \Bigl\langle A_{x+iy}(k) \Bigr\rangle_{1^+ \cdots n^+} \; \rlap{\ \
      $=$}\raisebox{1em}{tree} \; 0 \,, \\
    \Bigl\langle A_{x- i y}(k) \Bigr\rangle_{1^+ \cdots n^+} \;
    \rlap{\ \ $=$}\raisebox{1em}{tree} \; -  i  \sqrt{2} \;
    k_{0-z} \; \Bigl\langle \phi(k) \Bigr\rangle_{1 \cdots n} \,, &
    \Bigl\langle A_{0-z}(k) \Bigr\rangle_{1^+ \cdots n^+} \; \rlap{\ \
      $=$}\raisebox{1em}{tree} \; 0 \,.
  \end{array}
\end{eqnarray}
[compare with the self-dual {\it Ansatz}~(\ref{A-Phi-a})].

It is worth mentioning that the discussion extends to tree amplitudes
with two quark lines.\cite{bardeen} A right-handed quark current
$\Psi^{\dot\alpha}(1,\ldots,n;p)$ [resp.  left-handed antiquark
current $\bar\Psi^\alpha(q;1,\ldots,n)$] consists of $n$ on-shell
positive helicity gluons, a right-handed quark with on-shell momentum
$p$ [resp. a left-handed antiquark with on-shell momentum $q$] and an
off-shell antiquark [resp. an off-shell quark]. These currents verify
Berends-Giele type recurrence relations\cite{mahlon} which are
essentially a Dirac equation with a gauge field~(\ref{tree-id}).
Consider then:
\begin{eqnarray}
  &\rlap{\raisebox{1pt}{$/$}}\partial \Psi - \frac{ig}{\sqrt{2}}
  \rlap{$\,/$} A_{\rm SD} \Psi = 0 \,, \nonumber\\[-6pt]
  \\[-6pt]
  &\bar\Psi \rlap{\raisebox{10pt}{$\scriptstyle\leftarrow$}}
  \rlap{\raisebox{1pt}{$/$}}\partial - \frac{ig}{\sqrt{2}} \bar\Psi
  \rlap{$\,/$} A_{\rm SD} = 0  \,. \nonumber
\end{eqnarray}
With the representation~(\ref{A-Phi-a}) for the self-dual gauge field, the
right-handed quark current equation forces the Weyl spinor $\Psi$ to have the
form:
\begin{equation}
  \Psi = i \left( \begin{array}{c} \partial_{0-z} \\ - \partial_{x+iy}
  \end{array} \right) \Xi \,,
\end{equation}
where the scalar field $\Xi$ obeys the second order differential
equation:
\begin{equation}
  \label{Xi-equ}
  \partial^2 \Xi - ig \; \partial_{x+iy} \Phi \; \partial_{0-z} \Xi + ig
  \; \partial_{0-z} \Phi \; \partial_{x+iy} \Xi = 0 \,.
\end{equation}
Like $\Phi$, the field $\Xi$ is expanded in powers of $g$. For the the
self-dual solution $\Phi_{J(p) = p^2 j(p)}$, one finds:
\begin{eqnarray}
 \lefteqn{
   \Xi = \sum_{m=1}^\infty  \Xi^{(m)} \,, \qquad \Xi^{(m)} \propto g^{m-1}
   \,, } \nonumber\\[6pt]
 \lefteqn{
  \Xi^{(m)}(k) = ( ig )^{m-1} \int \frac{d^4p_1}{(2\pi)^4} \cdots
    \frac{d^4p_m}{(2\pi)^4} (2\pi)^4\delta(p_1 + \cdots + p_m - k)}
  \\
  &\times j(p_1) \cdots j(p_{m-1}) \; \Xi^{(0)}(p_m) \;
    (Q_1 - Q_2)^{-1} (Q_2 - Q_3)^{-1} \cdots (Q_{m-1} - Q_m)^{-1}
   \,, \nonumber\\[6pt]
  \lefteqn{
    p^2 \Xi^{(0)}(p) = 0 \,.} \nonumber
\end{eqnarray}
With the plane wave {\it Ansatz}~(\ref{planew}) for $j(p)$ and the
choice $\Xi^{(0)}(p) = \sqrt{|p_{0+z}|} / p_{x+iy}$, one recovers the
MHV quark current $\Psi^{\dot\alpha}(1,\ldots,n;p)$ of Mahlon {\it et
  al.}\cite{mahlon}

Similarly each of the two components $\bar\Psi^\alpha$ of the
left-handed antiquark current verifies the same second order
differential equation:
\begin{equation}
  \label{Psi-equ}
  \partial^2 \bar\Psi^\alpha - ig \; \partial_{x+iy} \Phi \; \partial_{0-z}
  \bar\Psi^\alpha + ig \; \partial_{0-z} \Phi \; \partial_{x+iy}
  \bar\Psi^\alpha = 0 \,,
\end{equation}
with solution:
\begin{eqnarray}
 \lefteqn{
   \bar\Psi = \sum_{m=1}^\infty \bar\Psi^{(m)} \,, \qquad \bar\Psi^{(m)}
   \propto g^{m-1} \,, } \nonumber\\[6pt]
 \lefteqn{
  \bar\Psi^{(m)}(k) = ( ig )^{m-1} \int \frac{d^4p_1}{(2\pi)^4} \cdots
    \frac{d^4p_m}{(2\pi)^4} (2\pi)^4\delta(p_1 + \cdots + p_m - k) }
  \\
  &\times \bar\Psi^{(0)}(p_1) \; j(p_2) \cdots j(p_m) \;
    (Q_1 - Q_2)^{-1} (Q_2 - Q_3)^{-1} \cdots (Q_{m-1} - Q_m)^{-1} \,,
   \nonumber\\[6pt]
  \lefteqn{
    \bar\Psi^{(0)}(q) = \left( \begin{array}{c} -q_{x+iy} \\ q_{0+z}
   \end{array} \right) g(q) \,, \qquad q^2 g(q) = 0 \,. } \nonumber
\end{eqnarray}
For $g(q) = 1 / \sqrt{|q_{0+z}|}$, the QCD antiquark
current\cite{mahlon} $\bar\Psi^\alpha(q;1,\ldots,n)$ is also recovered.

It is straightforward to find an action of two scalar fields
$\bar\psi,\xi$ that admits Eqs.~(\ref{Xi-equ}) and~(\ref{Psi-equ}) as
equations of motion:
\begin{equation}
  S_{\rm matter}(\Phi,\bar\psi,\xi) = \int d^4x \; \frac{1}{2} \bar\psi \left(
  \partial^2 \xi - ig \; \partial_{x+iy} \Phi \; \partial_{0-z} \xi + ig
  \; \partial_{0-z} \Phi \; \partial_{x+iy} \xi \right) \,,
\end{equation}
where $\Phi$ is the self-dual field~(\ref{equPhi}). If $\Phi$ is to be
dynamical, one can consider $S_{SC}(\phi) + S_{\rm
  matter}(\phi,\bar\psi,\xi)$.  Notice that the coupling to the two
other actions would modify the equation of motion for $\phi$. It is
also possible to extend the (classical) SDYM
symmetries~(\ref{equLambda}).  Eq.~(\ref{Xi-equ}) is invariant under
$\Xi \to \Xi + \Delta$ if the shift $\Delta$ satisfies:
\begin{equation}
  \partial^2 \Delta - ig \partial_{x+iy} \Phi \partial_{0-z} \Delta + ig
  \partial_{0-z} \Phi \partial_{x+iy} \Delta - \frac{ig}{f^2_\pi}
  \partial_{x+iy} \Lambda \partial_{0-z} \Xi + \frac{ig}{f^2_\pi}
\partial_{0-z}
  \Lambda \partial_{x+iy} \Xi = 0 \,.
\end{equation}
This equation is solved by a one-parameter family of symmetry
generators $\Delta_s$,
\begin{eqnarray}
  &\partial_{0-z} \Delta_{s+1} = \partial_{x-iy} \Delta_s -
  ig \; \partial_{0-z} \Phi \; \Delta_s + \frac{ig}{f^2_\pi} \;
  \Lambda_s \; \partial_{0-z} \Xi \,, \nonumber\\[-6pt]
  \\[-6pt]
  &\partial_{x+iy} \Delta_{s+1} = \partial_{0+z} \Delta_s -
  ig \; \partial_{x+iy} \Phi \; \Delta_s + \frac{ig}{f^2_\pi} \; \Lambda_s
  \; \partial_{x+iy} \Xi \,. \nonumber
\end{eqnarray}
which complete the previous $\Lambda_s$ given in~(\ref{Lambdas}).

\section{One-Loop Amplitudes}
\noindent
To go beyond tree level, one needs to consider the full actions. I
will only discuss the gauge sector and thus do not include the matter
action introduced at the end of the last section. Since the three SDYM
actions are different, their $S$-matrix are different too. We will see
in this section that their one-loop amplitudes are nevertheless equal
and moreover coincide with one-loop MHV amplitudes.

I use the following construction of a one-loop amplitude. Take a
truncated tree Green function with all legs on-shell except two which
are then sewn together with a propagator. Each insertion on the loop
corresponds to the multiplication by a factor:
\begin{eqnarray}
  - \frac{1}{f^2_\pi} \;
    \frac{\delta^2 S_{\rm DNS}(H)}{H^{-1} \delta H(k_1) \, H^{-1}
    \delta H(k_2)} \biggr|_{\rm classical\ solution} \,, \label{effacta}\\
  \frac{1}{f^2_\pi} \;
    \frac{\delta^2 S_{\rm LMP}(\phi)}{\delta\phi(k_1) \, \delta\phi(k_2)}
    \biggr|_{\rm classical\ solution} \,,
    \label{effactb}\\
  \left. \left( \begin{array}{cc}
    \frac{\delta^2 S_{\rm SC}(\phi,\Lambda)}{\delta\phi(k_1) \,
    \delta\phi(k_2)} &
    \frac{\delta^2 S_{\rm SC}(\phi,\Lambda)}{\delta\phi(k_1) \,
    \delta\Lambda(k_2)} \\
  \frac{\delta^2 S_{\rm SC}(\phi,\Lambda)}{\delta\Lambda(k_1) \,
    \delta\phi(k_2)} & 0 \end{array} \right)
    \right|_{\rm classical\ solutions} \,. \label{effactc}
\end{eqnarray}
This amounts to compare the one-loop effective actions of the three
models.\cite{gordon} Taking into account that:
\begin{eqnarray}
  H^{-1} \partial_{0+z} H &=& - ig \; \partial_{x+iy} \Phi
  \,, \nonumber\\[-6pt]
  \\[-6pt]
  H^{-1} \partial_{x-iy} H &=& - ig \; \partial_{0-z} \Phi
  \,, \nonumber
\end{eqnarray}
one verifies that~(\ref{effacta}), (\ref{effactb}) and~(\ref{effactc})
indeed match, up to a factor of two for $S_{\rm SC}$ (it has twice as
many fields that the other actions), and thus establishes the
equivalence of the three SDYM models at one-loop order.

Before comparing these results to QCD, I would like to indicate the
analogy between the generators of the (classical) SDYM symmetries and
the tree currents with two off-shell legs that we just used. The
latter are generated by the two-point off-diagonal matrix element:
\begin{eqnarray}
  \lefteqn{
  \langle0| \hat\phi(k) \hat\phi(q) |\rlap{\~\ } j\rangle^{\rm tree}_c =
  \sum_{m=1}^\infty \frac{i^m}{m!} \int \frac{d^4p_1}{(2\pi)^4} \cdots
  \frac{d^4p_m}{(2\pi)^4} } \qquad\qquad\qquad\quad \nonumber\\
  &&{} \times p_1^2 \, j(-p_1) \cdots p_m^2 \, j(-p_m) \; \langle
  \phi(p_1) \cdots \phi(p_m) \phi(k) \phi(q) \rangle_c^{\rm tree}
  \,. \nonumber\\
  \lefteqn{= \frac{\delta\Phi_J(k)}{i\delta J(-q))} \biggr|_{J(p) =
  p^2 j(p)} \,. }
\end{eqnarray}
One observes that $\delta \Phi_J(k) / i\delta J(-q)$ obeys the same
equation~(\ref{equLambda}) than the symmetry generators $\Lambda$.
Notice moreover that the solution to these equations
coincides\cite{daniel} with the generalized QCD current defined by
Mahlon {\it et al.}\cite{mahlon,daniel}

What about the MHV one-loop amplitudes as derived from QCD? In
principle one should compute the one-loop effective action for QCD.
However the comparison with the SDYM effective actions is difficult
owing to the tensor structure that appears in the loop. It is much
easier to use the one-loop equivalence between MHV QCD and scalar
QCD.\cite{tasi,daniel} The MHV one-loop
amplitudes~(\ref{one-loop}) can be derived from the action:
\begin{equation}
  \label{action-QCD}
  S_{\rm QCD} = \int d^4x \; \left( - \frac{1}{4} \; {\rm tr}\;
  F_{\mu\nu} F^{\mu\nu} + \frac{1}{2} \; {\rm tr}\; D_\mu \chi D^\mu \chi
  \right) \,,
\end{equation}
by keeping the gauge field classical and letting the field $\chi$ run
in the loop. A simple verification, using the tree
result~(\ref{tree-id}), shows that the factor:
\begin{equation}
  \frac{\delta^2 S_{\rm QCD}}{\delta\chi(k_1) \, \delta\chi(k_2)}
  \biggr|_{\rm classical\ solutions}
\end{equation}
coincides with~(\ref{effacta}), (\ref{effactb}) and (\ref{effactc}).

\section{Conclusions}
\noindent
In this article I have reviewed how one-loop MHV amplitudes arise from
a SDYM system. In fact three SDYM models have been proposed. They
coincide up to one-loop but differ at higher order. It is believed
that none of them will reproduce higher order QCD amplitudes.
Nevertheless, one still thinks that they might provide a quicker way
to derive one-loop results. This hope is based on the observation that
the symmetries of the SDYM equations play a role in the derivation of
one-loop SDYM amplitudes. According to Bardeen
suggestion,\cite{bardeen} the form of one-loop amplitudes might be a
consequence of an anomaly in these symmetries. In the verification of
this conjecture, one needs to better understand how to lift the
classical SDYM symmetries to the quantum theory.  Moreover it is not
clear whether the existence of an anomaly is compatible with a zero
beta function in the DNS model. This latter model has also been
proposed to describe the low energy matter sector of the $N=2$
heterotic string.\cite{ooguri} The $N=2$ heterotic string has been
shown to have vanishing amplitudes at all orders except the
three-point amplitude. The apparent inconsistency with the
non-vanishing one-loop amplitudes described here leads me to the
following two remarks. First, the spacetime signature is here $(3+1)$
whereas in the $N=2$ string it is $(2+2)$. Second, the low energy
action of the $N=2$ heterotic string also possesses a self-dual
gravity sector\footnote{Self-dual gravity can presumably be connected
  in the some way to MHV gravity amplitudes.}\ \ which couples non
trivially to the matter sector. It is not impossible that the sum of
the matter contribution, the gravity contribution as well as the
contribution coming from the mixing of the two add up to zero. Coming
back to QCD, we have seen that it is possible to include two quarks
lines in the amplitudes at the classical level and probably also at
one-loop order.

\nonumsection{Acknowledgements}
\noindent
I would like to thank Zvi Bern, Lance Dixon, Andrew Morgan and
V.P.~Nair for many interesting discussions and Michael Duff for
pointing out Ref.~1 to me. I also thank the organizers of the workshop
on Low Dimensional Field Theory in Telluride 1996 where part of this
work has been done. This work was partially supported by the NSF under
contract PHY-92-18990.

\nonumsection{References}

\end{document}